\documentclass[twoside,a4paper,11pt]{sea10}
\usepackage{graphicx}
\usepackage{hyperref}
\usepackage{movie15}
\usepackage[latin1]{inputenc}

\DeclareRobustCommand{\ion}[2]{%
\relax\ifmmode
\ifx\testbx\f@series
{\mathbf{#1\,\mathsc{#2}}}\else
{\mathrm{#1\,\mathsc{#2}}}\fi
\else\textup{#1\,{\mdseries\textsc{#2}}}%
\fi}

\newcommand{\hh}{\ion{H}{ii}~}

\newcommand{\ha}{H$\alpha$}

\newcommand{\mz}{{\small $\mathcal{M}$-Z}}
\newcommand{\mew}{{\small $\mathcal{M}$-EW(H$\alpha$)}}
\newcommand{\mze}{{\small $\mathcal{M}$-Z-EW(H$\alpha$)}}
\newcommand{\ewha}{$|$EW(\ha)$|$}

\begin{document}
\pagenumbering{arabic}
\pagestyle{myheadings}
\thispagestyle{empty}
{\flushleft\includegraphics[width=\textwidth,bb=58 650 590 680]{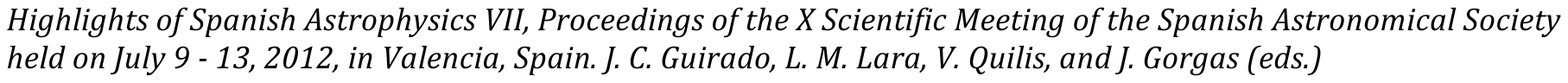}}
\vspace*{0.2cm}
\begin{flushleft}
{\bf {\LARGE
%
The oxygen abundance in the IFU era 
%
}\\
\vspace*{1cm}
%
F. F. Rosales-Ortega$^{1,2}$,
S. F. Sánchez$^{2}$, 
and 
A. I. Díaz$^{1}$
%
}\\
\vspace*{0.5cm}
%
$^{1}$
Departamento de F{\'i}sica Te{\'o}rica, Universidad Aut\'onoma de Madrid,
28049 Madrid, Spain.\\
$^{2}$
Centro Astron{\'o}mico Hispano Alem{\'a}n, Calar Alto, CSIC-MPG, C/Jes{\'u}s
Durb{\'a}n Rem{\'o}n 2-2, E-04004 Almeria, Spain.\\
%
\end{flushleft}
%
\markboth{
The oxygen abundance in the IFU era 
}{ 
%
Rosales-Ortega et al.
%
}
\thispagestyle{empty}
\vspace*{0.4cm}
\begin{minipage}[l]{0.09\textwidth}
\ 
\end{minipage}
\begin{minipage}[r]{0.9\textwidth}
\vspace{1cm}
\section*{Abstract}{\small
%
Spatially-resolved information of gas-phase emission provided by integral field
units (IFUs) are allowing us to perform a new generation of emission-line
surveys, based on large samples of \hh regions and full two-dimensional coverage.
Here we present two highlights of our current studies employing this
technique: 1) A statistical approach to the abundance
gradients of spiral galaxies, which indicates an {\em universal} radial
gradient for oxygen abundance; and 2) The discovery of a new scaling relation
of \hh regions in spiral galaxies, the {\em local} mass-metallicity relation
of star-forming galaxies.
\normalsize}
\end{minipage}
%
%
%
\section{Introduction}

Nebular emission lines from bright-individual \hh regions have been,
historically, the main tool at our disposal for the direct measurement
of the gas-phase abundance at discrete spatial positions in low
redshift galaxies. 
However, up to now most of the observations targeting nebular emission have
been made with single-aperture or long-slit spectrographs, resulting in a
small number of galaxies studied in detail, a small number of \hh regions
studied per galaxy, and a limiting coverage of these regions within the galaxy
surface. The advent of multi-object spectrometers (MOS) and integral field
spectroscopy (IFS) instruments with large fields-of-view (FoV) now offers us the
opportunity to undertake a new generation of emission-line surveys, based on
samples of scores to hundreds of \hh regions and full two-dimensional (2D)
coverage of the discs of nearby spiral galaxies. In the last few years we
started a major observational programme aimed to study the 2D properties of
the ionized gas and \hh regions in a representative sample of nearby face-on
spiral galaxies using IFS. We have catalogued more than $\approx2500$ \hh
regions with good spectroscopic quality in 38 galaxies; to our knowledge, this
is by far the largest 2D spectroscopic \hh region survey ever accomplished.

The spatially-resolved information provided by these observations are allowing us
to test and extend the previous body of results from small-sample studies,
while at the same time open up a new frontier of studying the 2D oxygen
abundance on discs and the intrinsic dispersion in metallicity, progressing
from a one-dimensional study (radial abundance gradients) to a 2D
understanding, allowing us at the same time to strengthen the diagnostic
methods that are used to measure \hh region abundances in galaxies.
Here we present two highlights of our current studies employing this large
spectroscopic database: 1) An IFS, statistical approach to the abundance
gradients of spiral galaxies; and 2) The discovery of a new scaling relation
of \hh regions in spiral galaxies and how we use it to to reproduce --with
{\bf remarkable} agreement-- the mass-metallicity relation of star-forming
galaxies.

\begin{figure}
\center
\includegraphics[height=6cm]{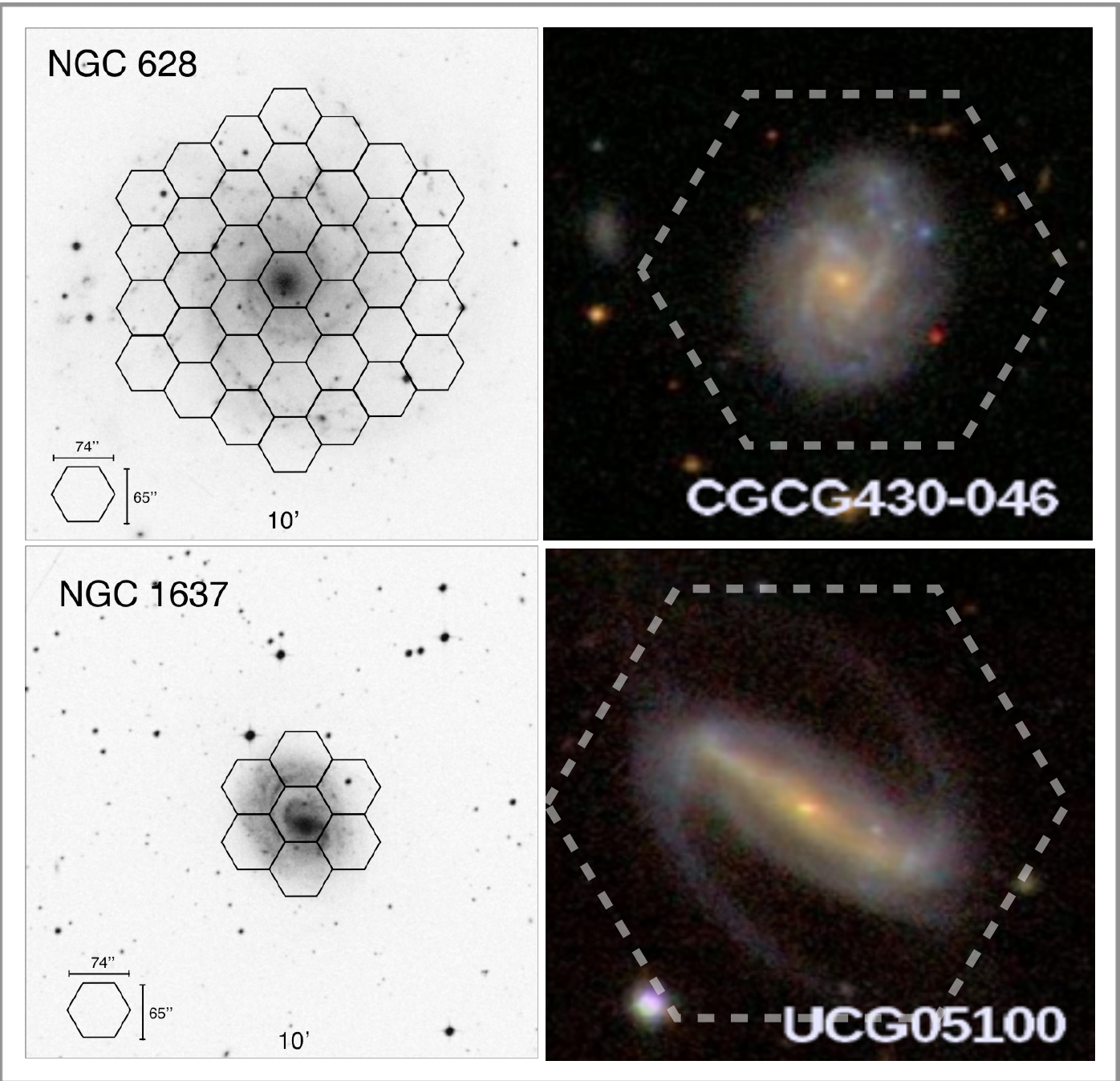} ~
\includegraphics[height=5.9cm]{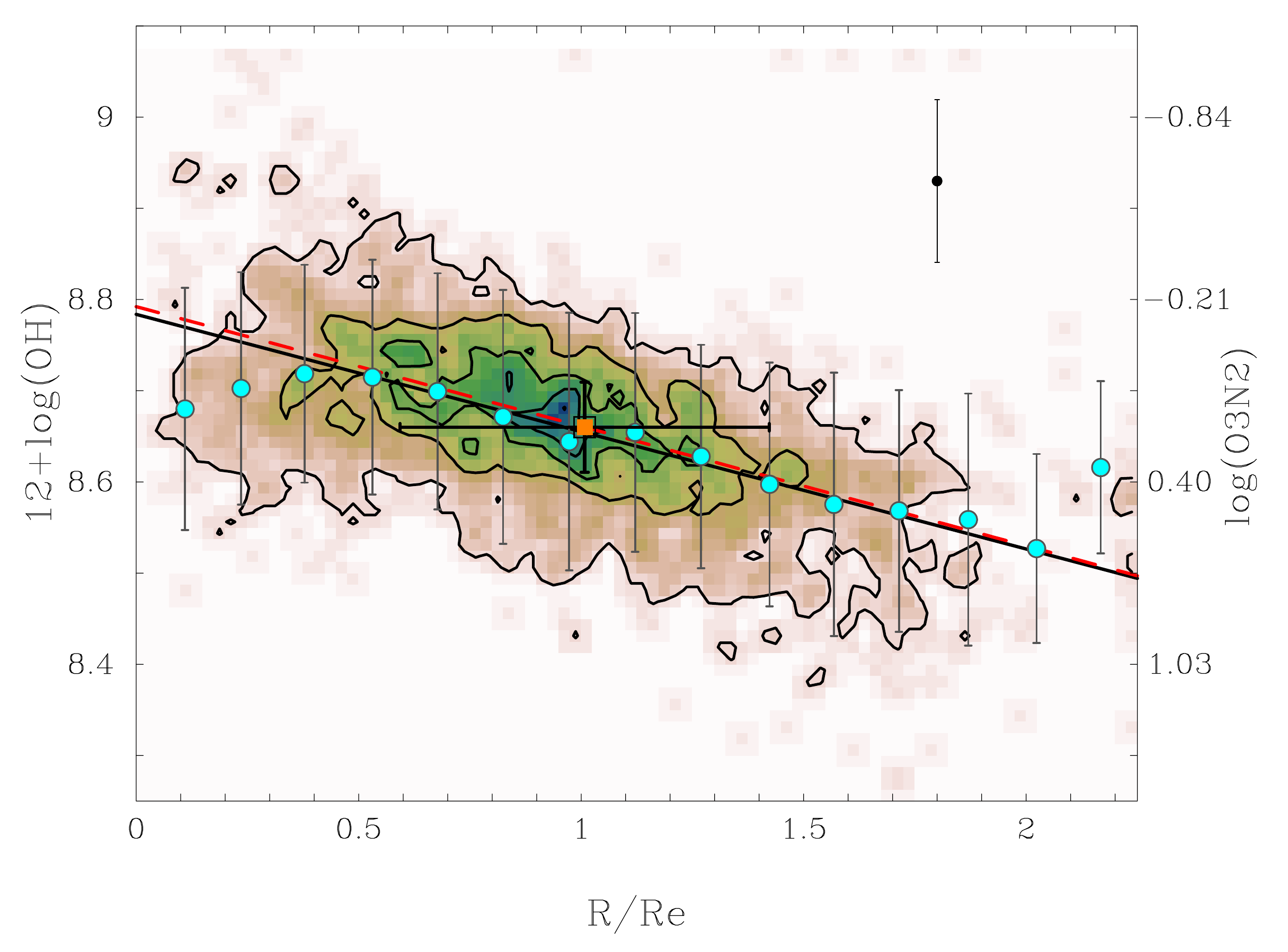}
\caption{\label{fig2} \small
  {\em Left:} Example of objects included in the sample, the overlaid hexagonal fields
  correspond to the PPAK instrument FoV.
  {\em Right:} Radial oxygen abundance density distribution for the whole \hh
  region spectroscopic sample.
  The first contour indicates the mean density, with a regular spacing
  of four times this value for each consecutive contour. The light-blue
  solid-circles indicate the mean value (plus $1-\sigma$ errors) for each
  consecutive radial bin of $\sim$0.15 R$_e$.
  The average error of the derived oxygen abundance is shown by a single error
  bar located a the top-right side of the panel. The solid-orange square
  indicates the average abundance of the solar neighbourhood, at the distance
  of the Sun to the Milky-Way galactic center.
}
\end{figure}

\section{Data sample and analysis}

The studies here described were performed using IFS data of a sample of
nearby disc galaxies (left-panel of Fig.~\ref{fig2}), part belonging to the
PINGS survey \cite{RosalesOrtega:2010p3836}, and a sample of face-on spiral
galaxies from \cite{MarmolQueralto:2011p4103},
as part of the feasibility studies for the CALIFA survey \cite{Sanchez:2012p4186}.
The observations were designed to obtain continuous coverage spectra of the
whole surface of the galaxies. The final sample
comprises 38 objects, with a redshift range between $\sim$0.001-0.025. 
They were observed with the PMAS spectrograph \cite{Roth:2005p2463} 
in the PPAK mode \cite{Verheijen:2004p2481,Kelz:2006p3341} on the 3.5m
telescope in Calar Alto with similar setup, resolutions and integration
times, covering their optical extension up to $\sim$2.4 effective radii within
a wavelength range $\sim$3700-7000~\AA.
Details on the sample, observing strategy, setups, and data reduction can be
found in \cite{RosalesOrtega:2010p3836}, and \cite{MarmolQueralto:2011p4103}.

The \hh regions in these galaxies were detected, spatially
segregated, and spectrally extracted using {\sc HIIexplorer} \cite{Sanchez:2012b}.
We detected a total of 2573 \hh regions with good spectroscopic quality. This
is by far the {\em largest} spatially-resolved, nearby spectroscopic \hh region
survey ever accomplished.
The emission lines were decoupled from the underlying stellar population
using {\sc fit3d} \cite{Sanchez:2007p3299}, following a robust and
well-tested methodology \cite{RosalesOrtega:2010p3836,Sanchez:2011p3844}.
Extinction-corrected, flux intensities of the stronger emission lines were
obtained and used to select only star-forming regions based on typical BPT
diagnostic diagrams.
The final sample comprises 1896 high-quality, spatially-resolved \hh
regions/aggregations of disc galaxies in the local Universe.
Details on the procedure can be found in \cite{Sanchez:2012b}.

\section{An IFS approach to abundance gradients}

We used our catalogue \hh regions to characterize the radial gradients of the
gas-phase physical properties of the galaxy sample. The radial distance was
normalised to the effective radius of each galaxy.
The right-panel of Fig.~\ref{fig2} shows the radial density distribution for the oxygen
abundance derived using the O3N2 indicator \cite{Pettini:2004p315}, once
scaled to the average value at the effective radius for each galaxy.
The solid line shows the average linear regression found for each individual
galaxy. The red dashed line shows the actual regression found for all the \hh
regions detected for all the galaxies.

As discussed in \cite{Sanchez:2012b}, our results seem to indicate that there
is an {\em universal} radial gradient for oxygen abundance when normalized
with the {\em effective radii} of the galaxies. For each galaxy we derived the
correlation coefficient, the slope and zero-point of a linear regression. A
histogram of the slopes of the gradients are consistent with a Gaussian
distribution, i.e. the dispersion of values found for each individual galaxy
is compatible with the average one, not showing strong statistical
deviations. A similar behaviour was found with the radial distribution of the
equivalent width of H$\alpha$.

\section{The local \mz\ relation}

\begin{figure}
\center
\includegraphics[height=5.5cm]{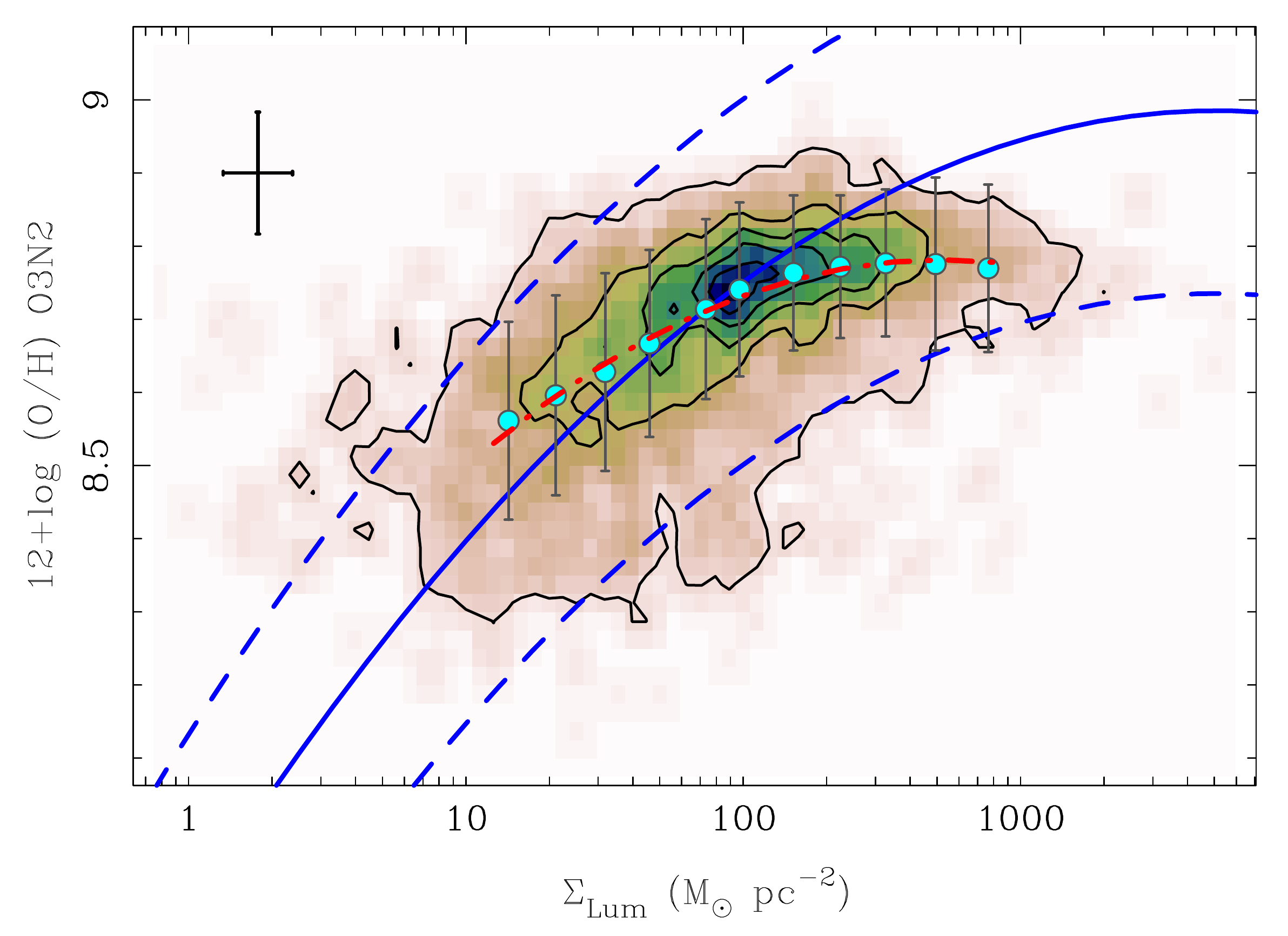} ~
\includegraphics[height=5.5cm]{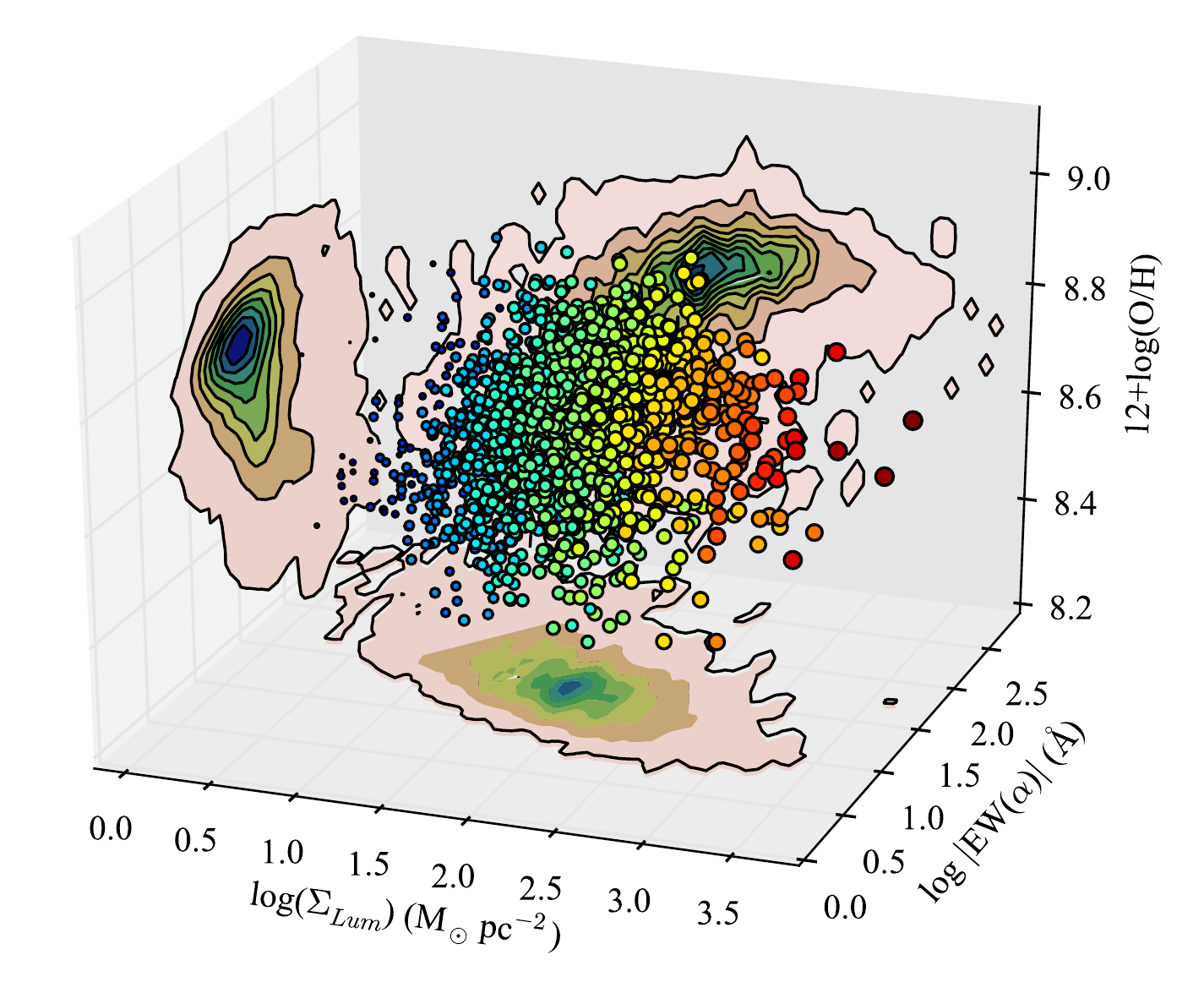} ~
\caption{\label{fig3} \small
  {\em Left-panel}: The relation between surface mass density and
  gas-phase oxygen metallicity for $\sim$2000 \hh regions in nearby
  galaxies, the {\em local} \mz\ relation.
  The first contour stands for the mean density value, with a regular
  spacing of four time this value for each consecutive contour. The blue
  circles represent the mean (plus 1$\sigma$ error bars) in bins of
  0.15\,dex. The red dashed-dotted line is a polynomial fit to
  the data. The blue-lines correspond to the \cite{Tremonti:2004p1138}
  relation ($\pm$0.2\,dex) scaled to the relevant units. Typical errors for
  $\Sigma_{\rm Lum}$ and metallicity are represented.
  {\em Right-panel}: 3D representation of the local $\mathcal{M}$-Z-EW(H$\alpha$) relation. 
    The size and color scaling of the data points are linked to the value of
    $\log \Sigma_{\rm Lum}$ (i.e. low-blue to high-red values). 
    The projection of the data over any pair of axes reduces to the
    local \mz, \mew, and metallicity-EW(H$\alpha$) relations.
    An online 3D animated version is available at: \url{http://tinyurl.com/local-MZ-relation}
}
\end{figure}

The existence of a strong correlation between stellar mass and gas-phase
metallicity in galaxies is a well known fact. The mass-metallicity (\mz)
relation is consistent with more massive galaxies being more metal-enriched,
it was established observationally by \cite{Tremonti:2004p1138}.
However, there has been no major effort to test the \mz\ relation
using {\em spatially-resolved} information. We used our IFS observations in
order to test the distribution of mass and metals {\em within} the discs of
the galaxies.
We derived the (luminosity) surface mass density ($\Sigma_{\rm Lum}$,
M$_{\odot}$ pc$^{-2}$) within the area encompassed by our IFS-segmented \hh
regions, using the prescriptions given by \cite{Bell:2001p210} to convert
$B-V$ colors into a $B$-band mass-to-light ratio ($M/L$).

The left-panel of Fig.~\ref{fig3} shows the striking correlation between
the local surface mass density and gas metallicity for our sample of nearby
\hh regions, i.e. the {\em local} \mz\ relation, extending over $\sim$3 orders
of magnitude in $\Sigma_{\rm Lum}$ and a factor $\sim$8 in metallicity
\cite{Rosales:2012b}.
The notable similarity with the global \mz\
relation can be visually recognised with the aid of the blue-lines which stand
for the \cite{Tremonti:2004p1138} fit ($\pm$0.2\,dex) to the global \mz\
relation, shifted arbitrarily both in mass and metallicity to coincide
with the peak of the \hh region \mz\ distribution. Other abundance
calibrations were tested obtaining the same shape (and similar fit) of the
relation.

In addition, we find the existence of a more general relation between
mass surface density, metallicity, and the equivalent width of \ha, defined 
as the emission-line luminosity normalized to the adjacent continuum flux, 
i.e. a measure of the SFR per unit luminosity \cite{Kennicutt:1998p3370}.
This functional relation is evident in a 3D space with orthogonal
coordinate axes defined by these parameters, consistent with \ewha\
being inversely proportional to both $\Sigma_{\rm Lum}$ and metallicity, as
shown in Fig.~\ref{fig3}.
As discussed in \cite{Rosales:2012b}, we interpret the local \mze\ relation as
the combination of: i) the well-known relationships between both the mass and
metallicity with respect to the differential distributions of these parameters
found in typical disc galaxies, i.e. the {\em inside-out} growth; and ii) the
fact that more massive regions form stars faster (i.e. at higher SFRs), thus
earlier in cosmological times, which can be considered a local
{\em downsizing} effect, similar to the one observed in individual galaxies.

In order to test whether the global \mz\ relation observed by
\cite{Tremonti:2004p1138} using SDSS data is a reflection (aperture
effect) of the local \hh region mass-density vs. metallicity relation, we
perform the following exercise. We simulate a galaxy with typical $M_B$ and
$B-V$ values drawn from flat distributions in magnitude ($-$15\,to\,$-$23) and
colour ($\sim$0.4$-$1). A redshift is assumed for the mock galaxy, drawn from
a Gaussian distribution with mean $\sim$0.1 and $\sigma=0.05$, 
with a redshift cut $0.02 < z < 0.3$ in order to resemble the SDSS
\cite{Tremonti:2004p1138} distribution.
The mass of the galaxy is derived using the integrated $B$-band
magnitudes, $B-V$ colours and the average $M/L$ ratio following
\cite{Bell:2001p210}. The metallicity of the mock galaxy is derived using the
local \mz\ relation  within an aperture equal to the SDSS fiber (3 arcsec),
i.e. the metallicity that corresponds to the mass density surface at this
radius. The process is repeated over 10,000 times in order to obtain a
reliable distribution in the mass and metallicity of the mock galaxies.

\begin{figure}
\center
\includegraphics[height=5.5cm]{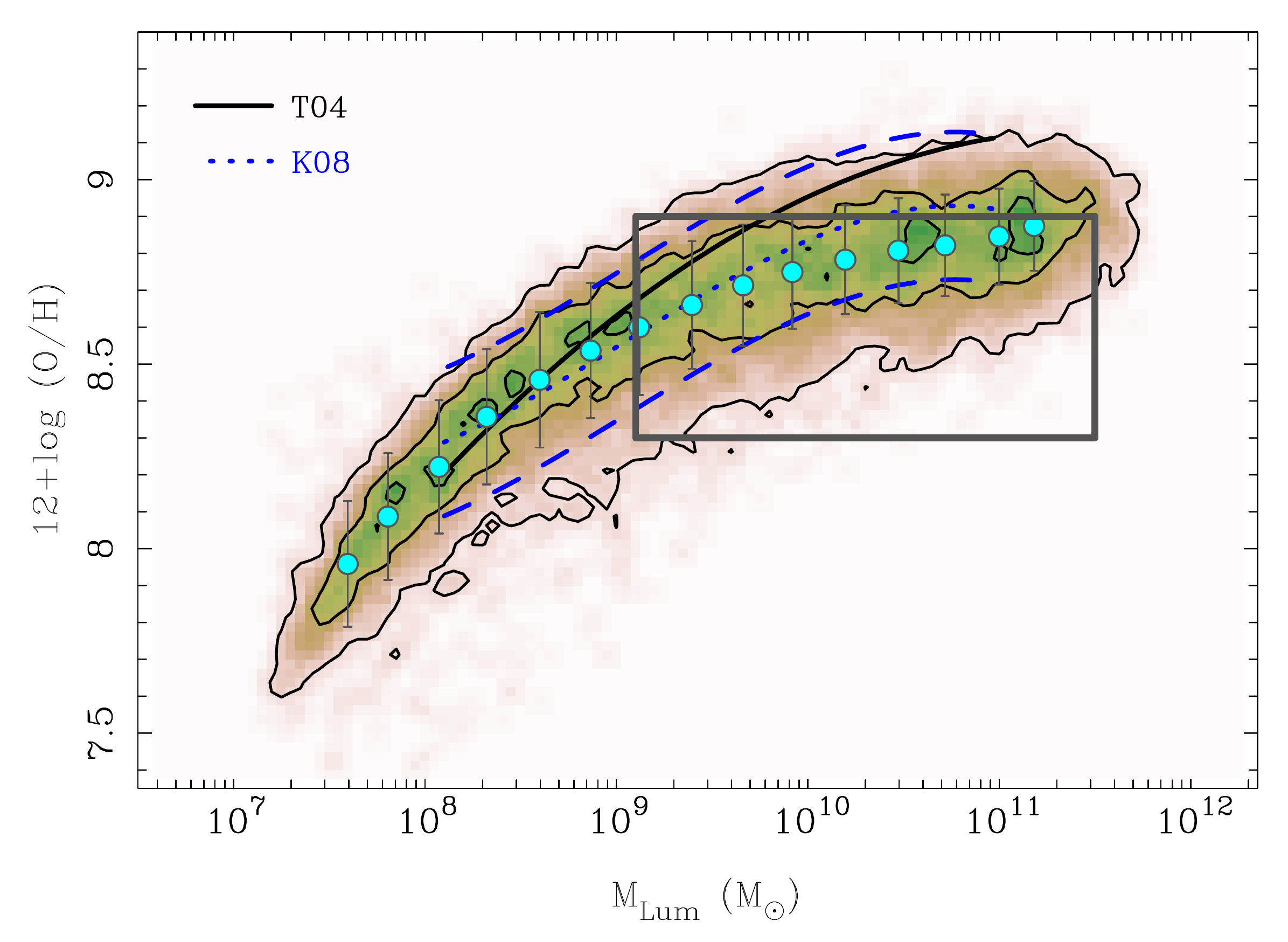}
\caption{\label{fig4} \small
  Distribution of simulated galaxies in the \mz\ plane assuming a 
  {\em local} \mz\ relation and considering the aperture effect of the SDSS
  fiber, as explained in the text.
  The contours correspond to the density of points, while the circles
  represent the mean value (plus 1$\sigma$ error bars) in bins of 0.15\,dex.
  The black line stands for the \cite{Tremonti:2004p1138} fitting,
  while the blue-lines correspond to the \cite{Kewley:2008p1394}
  $\pm$0.2\,dex relation.  
  The rectangle encompasses the range in mass and metallicity of the galaxy
  sample.
}
\end{figure}

Fig.~\ref{fig4} shows the result of the simulation, i.e. the
distribution of the mock galaxies in the \mz\ parameter space. We
reproduce --with a {\em remarkable} agreement-- the overall shape of the
global \mz\ relation assuming a local \mz\ relation and considering the
aperture effect of the SDSS fiber. The overlaid lines correspond to the
\cite{Tremonti:2004p1138} fit (black) and the \cite{Kewley:2008p1394}
$\pm$0.2\,dex relation (blue), for which the agreement is extremely good over
a wide range of masses.
The result is remarkable considering that we are able to reproduce the global \mz\
relation over a huge dynamical range, using a local \mz\ relation derived from
a galaxy sample with a restricted range in mass (9.2\,$<\log M_{\rm Lum}<$\,11.2)
and metallicity (8.3\,$<$\,12\,+log(O/H)\,$<$\,8.9), indicated by the rectangle
shown in Fig.~\ref{fig4}.

\section{Conclusions}

By using $\sim$2000 spatially-resolved \hh regions of a sample of nearby
galaxies observed with IFS, we found that the oxygen abundance presents a
radial gradient that, statistically, has the same slope for all the galaxies
when normalized to the effective radius, being independent of the morphology,
suggesting an {\em universal} radial gradient for oxygen abundance \cite{Sanchez:2012b}.
The same effect is also seen with the equivalent width of H$\alpha$. If true, this
result put strong constrains to the structure of spiral galaxies. 

We also demonstrate the existence of a {\em local} relation between the
surface mass density, gas-phase oxygen abundance and \ewha. The projection of
this distribution in the metallicity vs. $\Sigma_{\rm Lum}$ plane is the
{\em local} \mz\ relation, which notably has the same shape as the global \mz\
relation for galaxies.
We use the local \mz\ relation to reproduce --with an {\em outstanding}
agreement-- the global \mz\ relation by
means of a simple simulation which considers the aperture effects of the
SDSS fiber at different redshifts. We conclude that, the \mz\ relation in
galaxies is a scale-up integrated effect of a local \mz\ relation in the
distribution of star-forming regions across the discs of galaxies \cite{Rosales:2012b}.

\section*{Acknowledgments}   
{\small
Based on observations collected at the Centro
Astronómico Hispano-Alemán (CAHA) at Calar Alto, operated jointly by the
Max-Planck Institut für Astronomie and the Instituto de Astrofísica de
Andalucía (CSIC).
F.F.R.O. acknowledges the Mexican National Council for Science and Technology
(CONACYT) for financial support under the programme Estancias Posdoctorales y
Sabáticas al Extranjero para la Consolidación de Grupos de Investigación,
2010-2011 A.D. thanks the Spanish Plan Nacional de Astronomía programme
AYA2010-21887 C04-03.
}

%

%
\end{document}